 \documentclass[aps,prd]{revtex4-2}
 \usepackage{graphicx}
 \usepackage{amsmath}
  \newcommand{\VEC}[1]{\vec {#1} }
\begin{document}
\title{
Particle acceleration driven by null electromagnetic fields 
near a Kerr black hole}
\author{{Yasufumi Kojima}}
\email{ykojima-phys@hiroshima-u.ac.jp}
\author{Yuto Kimura}
 \affiliation{Department of Physics, 
Graduate School of Advanced Science and Engineering,
Hiroshima University, 
Higashi-Hiroshima, Hiroshima 739-8526, Japan}
\date{\today}
%

\begin{abstract}
Short timescale variability is often associated with a black hole system. The consequence of an electromagnetic outflow suddenly generated near a Kerr black hole is considered assuming that it is described by a solution of a force-free field with a null electric current. We compute charged particle acceleration induced by the burst field.
We show that the particle is instantaneously accelerated to the relativistic regime by the field with a very large amplitude, which is characterized by a dimensionless number $\kappa$.
Our numerical calculation demonstrates how the trajectory of the particle changes with $\kappa$.
We also show that the maximum energy increases with $\kappa^{2/3}$.
The typical maximum energy attained by a proton for an event near a super massive black hole is $E_{\rm max} \sim 100$ TeV, which is enough observed high-energy flares.
\end{abstract}


 \maketitle 
%

\section{Introduction}
Active galactic nuclei are very bright sources, and collimated relativistic 
plasma jets launched from the inner part are often observed in the universe.
Their activities are powered by accreting flow to supermassive 
black holes (SMBHs) that lurk at the center.
Recently, for the first time, the event horizon telescope (EHT) 
team \cite{2019ApJ...875L...1E} acquired 
a near-horizon image of a nearby galaxy M87, which is an example of an SMBH.
The resolved structure confirms what many researchers anticipated for many years.
Black hole astrophysics relevant to both observational and theoretical approaches are expected to be researched further.
For example, the central part of a black hole can be described using the Kerr metric with 
high precision, or using a new theory.
%

General relativistic magneto-hydro-dynamic (MHD) simulation is a powerful tool to study the nature near a black hole theoretically.
Several studies have clearly demonstrated the site of conversion 
from material inflow to outward jet launching. 
However, as a different approach,
the so-called force-free electrodynamics (FFE) is
employed for tenuous plasma regions such as those in the magnetosphere 
and in the magnetized jet.
In the approach, material motions are decoupled, so that 
only the electromagnetic field can be solved under certain conditions.
The Blandford--Znajek process\cite{1977MNRAS.179..433B} of
extraction of energy from a rotating black hole 
was studied assuming a steady force-free magnetosphere.
The magnetosphere is governed by the highly nonlinear equations of FFE, and the analytic solutions are limited; 
the simplest being a split-monopole solution\cite{1973ApJ...180L.133M}.
It describes a radial magnetic field near the center and a 
rotation-induced outward electromagnetic field at infinity. 
The radial field generated by the magnetic monopole charge is unphysical; however, the solution is useful to study the behavior away from the central object. Therefore, the global structure of the magnetosphere is 
studied using a perturbation method from the solution in the Schwarzschild spacetime\cite{2008PhRvD..78b4004T,2015PhRvD..91f4067P,2015ApJ...812...57P,2017ApJ...836..193P} (a problem regarding the perturbation method
is discussed in \cite{2018PhRvD..98h4056G}). Other approaches require intense numerical computations; for example, solving the Grad--Shafranov equation with some iteration techniques 
\cite{2004ApJ...603..652U,2005ApJ...620..889U,2014ApJ...788..186N,
2018MNRAS.477.3927M} and performing the time-dependent calculations\cite{
2004MNRAS.350..427K,2004MNRAS.350.1431K,2006MNRAS.367.1797M,
2010PhRvD..82d4045P,2013PhRvD..88j4031P,2016MNRAS.455.3779P,
2017PhRvD..96f3006C,2017CQGra..34u5001E}.
%

Recently, a class of exact solutions satisfying the force-free condition in a Kerr spacetime were derived\cite{2013CQGra..30s5012B,2014MNRAS.445.2500G}
(see also early works \cite{2005ApJ...635.1197M,2011MNRAS.417.1098M}).
The solution was reduced to a split-monopole solution as a limiting case. The solutions were characterized by the fact that two Lorentz invariant scalars vanish: ${\VEC E}\cdot {\VEC B}=0$ and $B^2-E^2=0$.
Such an electromagnetic field propagates at the speed of light, and the four-current is a null vector. The field is analogous to the traveling-wave mode that appears in the dynamical perturbations of 
stationary force-free solutions \cite{
2014PhRvD..90j4022Y,2015PhRvD..91l4055Y,2015PhRvD..92b4049Z}.
The wave propagates along the principal null direction of the background spacetime without any back scattering. An attempt to resolve the null four-current was discussed\cite{2015PhRvD..92b4054M,2015GReGr..47...52M} and it involved artificial decomposition to a linear combination of two time-like four-currents with opposite charges.
%

It is unclear how such a null field is produced in astrophysical scenarios. 
Some theoretical works however suggest a possibility occurred near a black hole horizon.
It is shown that any stationary, axisymmetric and regular FFE solution reduces to the same form in the near-horizon limit of extreme Kerr \cite{2016PhRvD..93j4041G,2020JCAP...10..048C}.
The “attractor solution” is still stationary and axisymmetric, but it is null.
The near-horizon region of highly spinning black hole is very important for astrophysical scenarios such as the jet formation.
Magnetically dominated solutions $ (B^2-E^2>0) $ are stable, whereas electrically dominated solutions $ (B^2-E^2<0) $ are unstable since the electromagnetic energy is transferred to plasma acceleration and/or heating.
Some energy may be ejected as the null field.

Present numerical codes cannot simulate the situation, but some results are suggestive.
Through the numerical calculations of time-dependent FFE, it is well known that current sheets may develop within the black hole magnetospheres \cite{2004MNRAS.350..427K,2010PhRvD..82d4045P,
2017PhRvD..96f3006C,2018PhRvD..98b3008E}. 
For asymptotically uniform magnetic configuration, the singular current-sheet forms on the equator inside ergo-sphere of a rapidly rotating black hole.
As described in \cite{2004MNRAS.350..427K}, part of the outward energy and angular momentum can be traced to the current sheet.
The equatorial boundary condition is therefore important to determine the global solution \cite{2018PhRvD..98b3008E,2018PhRvD..98d3023P}. 
Numerical simulations show that magnetic dominance ($B^2 -E^2>0$) tends to break down in the singular regions.
The calculation requires artificial reduction of the electric field to ensure $B^2-E^2 \approx 0$ \cite{2004MNRAS.350..427K,2010PhRvD..82d4045P,
2017PhRvD..96f3006C,2018PhRvD..98b3008E}.
This mimics some physical mechanisms of electromagnetic energy loss; the energy is locally dissipated.
Some part of the energy is also carried away by a burst wave, which may be approximated by a null field, as suggested by the perturbation analysis of FFE  \cite{
2014PhRvD..90j4022Y,2015PhRvD..91l4055Y,2015PhRvD..92b4049Z}.
Thus, we expect that the null FFE field or the field approximated by it must appear as a burst near a black hole and propagate outwardly. We consider the consequence of the burst field approximated by the null FFE solution. 
We discuss plasma motion under strong gravity when the electromagnetic burst-field passes. The charged particle is treated as a test particle, and its interaction with the incident null field is studied.
The rest of this manuscript is organized as follows. In section 2, we review the null FFE solutions in Kerr spacetime. The original work\cite{2013CQGra..30s5012B} was based on the Newman--Penrose formalism\cite{1962JMP.....3..566N}. Their result is summarized using the 3 + 1 formalism for complemental understanding. In section 3, we explicitly present axially symmetric magnetosphere relevant to some astrophysical situations. 
In section 4, we calculate a charged particle motion driven by the electromagnetic field in the curved spacetime. The particle moves by the electromagnetic acceleration and gravitational attraction. 
The motion is trivial only in two extreme limits, for which one of forces exceeds the other in magnitude.  
We numerically estimate the critical value and examine the acceleration behavior by the electromagnetic burst. 
Finally, we discuss the results and provide the summary in Section 5.
We use geometrical units of $c=G=1$.

\section{Electromagnetic fields}
We briefly summarize the relevant equations in a Kerr spacetime with the same notations in \cite{1986bhmp.book.....T}. The metric in the Boyer--Lindquist coordinate is given by 
\begin{equation}
ds^2=-\alpha^2dt^2+\frac{\rho^2}{\Delta}dr^2
       +\rho^2d\theta^2 +\varpi^2(d\phi-\omega dt)^2,
\end{equation}
where
\begin{align}
&
\alpha^2=\frac{\rho^2\Delta}{\Sigma^2},
~~\rho^2=r^2+a^2\cos^2\theta,
~~\Delta=r^2+a^2-2Mr,
\nonumber
\\
&  \varpi^2=\frac{\Sigma^2}{\rho^2}\sin^2\theta ,
~~\omega=\frac{2Mar}{\Sigma^2},
~~\Sigma^2=(r^2+a^2)^2-a^2\Delta\sin^2\theta .
\end{align}

Electromagnetic vectors $\VEC{E}$ and $\VEC{B}$ refer to quantities measured in the fiducial observer (FIDO), which is sometimes called the 
nonrotating zero angular momentum observer (ZAMO). Using the vector analysis in a three-dimensional curved space, the Maxwell equations are expressed as

\begin{equation}
\VEC{\nabla}\cdot\VEC{E}=4\pi\rho_{e},
  \label{eqnMxw1}
\end{equation}

\begin{equation}
\VEC{\nabla}\cdot\VEC{B}=0,
  \label{eqnMxw2}
\end{equation}
\begin{equation}
\partial _{t} \VEC{B} +
\VEC{\nabla}\times( \alpha{\VEC{E}} + \VEC{\beta} \times \VEC{B} )=0,
  \label{eqnMxw3}
\end{equation}
\begin{equation}
- \partial _{t} \VEC{E} +
\VEC{\nabla}\times( \alpha{\VEC{B}} -\VEC{\beta} \times \VEC{E} )=
4\pi( \alpha\VEC{j}- \rho _{e} \VEC{\beta}),
  \label{eqnMxw4}
\end{equation}
where
\begin{equation}
  \VEC{\beta} = -\omega \varpi {\VEC e}_{{\hat \phi}}.
\end{equation}

In this paper, ${\hat i}$ denotes the component of a vector
in the orthogonal basis. Four-dimensional basis vectors are given by
\begin{equation}
{\VEC e}_{{\hat 0}}=\frac{1}{\alpha}(\partial_{t}-{\VEC \beta}),
~~{\VEC e}_{{\hat r}}=\frac{{\sqrt \Delta }}{\rho}\partial_{r},
~~{\VEC e}_{{\hat \theta}}=\frac{1}{\rho}\partial_{\theta},
~~{\VEC e}_{{\hat \phi}}=\frac{1}{\varpi}\partial_{\phi}.
\end{equation}

\subsection{Force-free electromagnetic field with null current}
We impose two conditions on the electromagnetic field
to construct a magnetosphere in a curved spacetime.
They include an ideal MHD condition
\begin{equation}
{\VEC E } \cdot {\VEC B}=0,
  \label{MHDcond}
\end{equation}
and a force-free condition 
\begin{equation}
\rho_{e}{\VEC E}+{\VEC j} \times {\VEC B}=0.
  \label{FFcond}
\end{equation}
Many studies have calculated electromagnetic fields under these conditions.
Numerical calculations are required except for the simplified scenarios in a flat spacetime. A class of exact FFE solutions in Kerr spacetime \cite{2013CQGra..30s5012B,2014MNRAS.445.2500G} is remarkable; they
yield an analytical solution when the four-electric current is proportional to the principal null congruence of the Kerr spacetime. Their solutions are derived by the Newman--Penrose formalism\cite{1962JMP.....3..566N}; however, the solutions are provided in terms of the 3+1 language for complemental understanding.
Three-dimensional current ${\VEC j}$ is expressed by charge density $\rho_{e}$ and a unit vector ${\VEC n}$ as  

\begin{equation}
{\VEC j} = \rho_{e}{\VEC n},
\label{curtrhnn}
\end{equation}
with
\begin{equation}
 {\VEC n}=\frac{\epsilon_{\rm out/in} \Sigma}{r^2+a^2}{\VEC e}_{{\hat r}}
+\frac{a \sqrt{\Delta} \sin \theta}{r^2+a^2}{\VEC e}_{{\hat \phi}},
\label{nullfm}
\end{equation}
where $\epsilon_{\rm out} =+1$ and $\epsilon_{\rm in} =-1$ for outgoing and incoming fields, respectively. The four-dimensional null vector with ${\VEC n}$ corresponds to outgoing and incoming principal null congruences. The condition (\ref{FFcond}) obtained using the current (\ref{curtrhnn}) is reduced to ${\VEC E }+{\VEC n} \times {\VEC B}=0$ in the case of $\rho_{e} \ne 0$.
The condition ${\VEC E}+{\VEC n} \times {\VEC B}=0$ leads to the orthogonality ${\VEC E} \cdot{\VEC B}=0$ in eq.~(\ref{MHDcond}). The null field condition $B^2 -E^2=0$ is obtained by assuming ${\VEC n}\cdot{\VEC B}=0$.
The null electromagnetic fields can be expressed by two functions provided below: $S_{\rm out/in}$ and $T_{\rm out/in}$. These functions correspond to the real and imaginary parts of one complex function in the Newman--Penrose formalism. The electromagnetic field components are explicitly written as
\begin{equation}
{\VEC B} =
  \frac{a}{\Sigma}S_{\rm out/in}{\VEC e}_{{\hat r}}
 -\frac{r^2+a^2}{\alpha \varpi \Sigma}T_{\rm out/in}{\VEC e}_{{\hat \theta}}
 -\frac{\epsilon_{\rm out/in}S_{\rm out/in}}{\alpha\varpi}{\VEC e}_{{\hat \phi}},
\label{BFFcomp}
\end{equation}

\begin{equation}
{\VEC E}
= -\frac{a}{\Sigma}T_{\rm out/in}{\VEC e}_{{\hat r}}
 -\frac{r^2+a^2}{\alpha \varpi \Sigma}S_{\rm out/in}{\VEC e}_{{\hat \theta}}
+ \frac{\epsilon_{\rm out/in}T_{\rm out/in}}{\alpha\varpi}{\VEC e}_{{\hat \phi}}.
\label{EFFcomp}
\end{equation}
The functions $S_{\rm out}$ and $T_{\rm out}$ are provided in the outgoing coordinates as $S_{\rm out} ( u, \theta, \chi)$ and $T_{\rm out} ( u, \theta, \chi)$, where 

\begin{equation}
u\equiv t -  \int \frac{r^2+a^2}{\Delta} dr,
~~
\chi \equiv \phi - \int \frac{a}{\Delta}dr .
\end{equation}
The ingoing fields are expressed using ingoing coordinates.
Ingoing solutions with $\epsilon_{\rm in}$ are obtained by $t \to - t$ and $\phi \to -\phi$.
Poynting flux at $r\gg M$ for these fields is given by
\begin{equation}
{\VEC S}_{\rm Poyn} = 
\frac{\epsilon_{\rm out/in}}{4\pi r^2 \sin^2 \theta}
\left( S_{\rm out/in} ^2 +T_{\rm out/in} ^2 \right){\VEC e} _{\hat{r}}.
\end{equation}
It is understood that $S_{\rm out/in}$ and $T_{\rm out/in}$ are associated with the relevant outgoing/ingoing flux.
Hereafter, we limit our consideration to the outgoing fields.
The Maxwell equations (\ref{eqnMxw1})-(\ref{eqnMxw4}) for (\ref{BFFcomp}) and (\ref{EFFcomp}) are reduced to a pair of first-order differential equations of $S_{\rm out}$ and $T_{\rm out}$ given as
\begin{equation}
a\partial_{u} S_{\rm out} + \frac{1}{\sin^2 \theta} \partial_{\chi} S_{\rm out}
+ \frac{1}{\sin \theta} \partial_{\theta} T_{\rm out} =0,
\label{BOUTcomp1}
\end{equation}

\begin{equation}
a \partial_{u} T_{\rm out} + \frac{1}{\sin^2 \theta} \partial_{\chi} T_{\rm out}
- \frac{1}{\sin \theta} \partial_{\theta} S_{\rm out} = 4\pi h_{\rm out},
\label{BOUTcomp2}
\end{equation}
where the source term $h_{\rm out}(u, \theta, \chi)$ is related to the charge density as
\begin{equation}
 \rho_{e} = \frac{r^2+a^2}{\alpha \Sigma^{2}} h_{\rm out} . 
\label{chargedn}
\end{equation}

Here, we summarize the mathematical method to calculate the null electromagnetic fields. Once two functions $S_{\rm out}$ and $T_{\rm out}$ are solved for the source term $h_{\rm out}$, which is related to the charge density and electric current, the electromagnetic fields can be expressed using these functions.
These functions $S_{\rm out}$ and $T_{\rm out}$ may be flexibly specified, since the initial configuration outwardly propagates with the light velocity.
In next section, we explicitly model the functions 
to apply an astrophysical problem.

\section{Axially symmetric magnetosphere}
We consider the following separated form of
$S_{\rm out} (u, \theta)$ and $T_{\rm out} (u, \theta)$,
\begin{equation}
 S_{\rm out}  = W(u) Q(\theta),
~~T_{\rm out}  = -a W^\prime (u) P(\theta),
  \label{STaxym}
\end{equation}
where $W=W(u)$ and the angular function 
$Q(\theta) \equiv (\sin \theta)^{-1}  P^\prime (\theta)$.
A prime denotes a derivative.
Equation (\ref{BOUTcomp1}) is automatically satisfied with this choice 
(\ref{STaxym}). Equation (\ref{BOUTcomp2}) is reduced to
\begin{equation}
-4\pi h_{\rm out} 
= a^2 W^{\prime \prime} P +W
(\sin \theta)^{-1}Q^\prime .
\end{equation}
The charge density is given by eq.(\ref{chargedn}), and the electric current ${\VEC j}$ is given by eq.(\ref{curtrhnn}).
The electromagnetic fields (\ref{BFFcomp}) and (\ref{EFFcomp})
are explicitly rewritten as

\begin{equation}
{\VEC B}= \frac{aW Q}{\Sigma}{\VEC e}_{{\hat r}}
 +\frac{a(r^2+a^2)W^{\prime} P}{\alpha \varpi \Sigma}{\VEC e}_{{\hat \theta}}
 -\frac{WQ}{\alpha\varpi}{\VEC e}_{{\hat \phi}},
\label{BfldM0}
\end{equation}

\begin{equation}
{\VEC E}= \frac{a^2 W^{\prime} P}{\Sigma}{\VEC e}_{{\hat r}}
 -\frac{(r^2+a^2)W Q}{\alpha \varpi \Sigma}{\VEC e}_{{\hat \theta}}
 -\frac{aW^{\prime} P}{\alpha\varpi}{\VEC e}_{{\hat \phi}} .
\label{EfldM0}
\end{equation}
Note that the terms containing $W^{\prime}$, ``time-derivative'' of $W$, vanish in the Schwarzschild limit. The amplitude of the electromagnetic field changes with time; however, the direction is always fixed in the spherically symmetric spacetime. The behavior in a Kerr case is more complicated.
Both the amplitude and direction of the fields vary with time. Some components vanish when the fields settle to a stationary state.
It is helpful to write down the potentials ${\VEC A}$ and $\Phi$ to derive these electromagnetic fields,
\begin{equation}
{\VEC B} = \VEC{\nabla} \times {\VEC A},
~~
\alpha {\VEC E}+{\VEC \beta} \times{\VEC B} = - \partial_{t} {\VEC A} -\VEC{\nabla} \Phi.
%
\end{equation}
Two examples are provided below.
\begin{align}
({\rm I}) &
~~[\Phi,{\VEC A}] = \left[0,
~-\frac{a^2PW}{\rho{\sqrt \Delta}} {\VEC e}_{{\hat r}}
+\frac{Q W^{\dagger}}{\rho \sin \theta} {\VEC e}_{{\hat \theta}}
+\frac{aW P}{\varpi} {\VEC e}_{{\hat \phi}}
\right],
%
\\
({\rm II}) &
~~[\Phi, {\VEC A}] =\left[W Q^{\dagger},
~\frac{W}{\rho{\sqrt \Delta}}((r^2+a^2)Q^{\dagger}-a^2P) {\VEC e}_{{\hat r}}
+\frac{aW P}{\varpi} {\VEC e}_{{\hat \phi}}
\right],
\end{align}
where
\begin{equation}
W^{\dagger}=\int^{u}Wdu^{\prime},
~~
Q^{\dagger}=\int^{\theta}(\sin \theta^{\prime})^{-1}Qd\theta^{\prime}.
\end{equation}
Note that the azimuthal component $A_{\phi}$ is gauge invariant for only the axially symmetric field.

\begin{figure}[t]
\centering
 \includegraphics[scale=0.75]{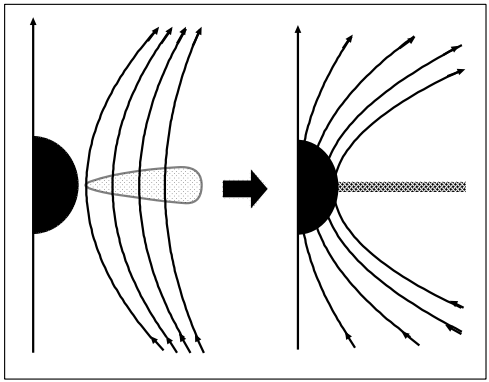}
\caption{ 
\label{Fig1}
The formation of the magnetosphere is described by the split-monopole solution near a black hole. Accreting matter on the equator drags the global magnetic field (left panel). Next, a split-monopole-like magnetosphere is formed (right panel).}
\end{figure}

%
The angular function $P(\theta)$ in eq.(\ref{STaxym}) is an arbitrary function. To specify this function, we consider the astrophysical possibility of a magnetosphere near a black hole.
The poloidal magnetic field far from the black hole is initially penetrated from $-z$ to $+z$ as shown in Fig.~\ref{Fig1}. The magnetic field is frozen to accreting matter on the equator. The rotation of the matter drags the magnetic field in the azimuthal direction. After the matter falls, the magnetic field formed outside it has a specific symmetry: both radial $B_{{\hat r}}$ and azimuthal components $B_{{\hat \phi}}$ change their direction across the equator. In the case of a sharp change in the magnetic field, a current 
sheet may remain on the equator. Based on this consideration, the function $Q(\theta)$ denotes an odd function with respect to $\theta=\pi/2$; $P(\theta)$ is an even function.
We first select the angular function $P(\theta)$ as 
\begin{equation}
 P= \frac{1}{3}(2+|\cos \theta |)(1-|\cos \theta |)^2 .
\label{selp0m}
\end{equation}
This function $P$ is continuous in $0\le \theta \le \pi$; however, its derivative contains a discontinuity at $\theta = \pi /2$. The function $Q(\theta)$ is given by
\begin{equation}
Q=(1-2H_{\pi/2})\sin^{2} \theta, 
\label{selq0m}
\end{equation}
where a function $H_{\pi/2}(\theta)$ is the unit step function $H_{\pi/2}=0 $ for $\theta <\pi/2$, and $H_{\pi/2}=1 $ for $\theta >\pi/2$. That is, $Q=\pm\sin^{2} \theta$. We explicitly write down the charge density
\begin{equation}
4 \pi \rho_{\rm e} =-\frac{r^2+a^2}{\alpha \Sigma^2}
\left[ a^2 W^{\prime \prime} P + 2W\left\{
(1-2H_{\pi/2})\cos\theta
 -\delta_{\rm D}(\theta-\frac{\pi}{2} ) \right\} 
\right],
\label{chrgM0}
\end{equation}
where $\delta_{\rm D}(=H_{\pi/2}^{\prime})$ is Dirac's delta function.
For a while, we consider the limit of $a = 0$. Equations (\ref{BfldM0})-(\ref{EfldM0}) are reduced to 
\begin{equation}
{\VEC B} = -\frac{W}{\alpha r} 
(1-2H_{\pi/2})\sin\theta {\VEC e}_{\hat \phi},
~~
{\VEC E} = -\frac{W}{\alpha r} 
(1-2H_{\pi/2})\sin\theta {\VEC e}_{\hat \theta}.
\label{montdBE}
\end{equation}
It is possible to add the magnetic monopole solution ${\VEC B} = \pm \mu_{M} r^{-2} {\VEC e}_{\hat r}$ to eq.~(\ref{montdBE}), where $\mu_{M}$ is a constant representing magnetic charge. The total electromagnetic field becomes a solution of FFE satisfying the magnetic dominance condition.

We explicitly express the field as ${\VEC B} = \mu_{M} r^{-2} {\VEC e}_{\hat r}$ in the upper hemisphere, ${\VEC B} =-\mu_{M} r^{-2} {\VEC e}_{\hat r}$ in the lower hemisphere, and $W=\mu_{M}\Omega$ in eq.~(\ref{montdBE}), where $\Omega$ denotes the angular velocity of the magnetic field line. The combined field  is the split-monopole solution\cite{1973ApJ...180L.133M}. 
It is possible to introduce the time-dependent function of $\Omega$ \cite{2011PhRvD..83l4035L}, that is, $W=\mu_{M}\Omega(u)$ in eq.(\ref{montdBE}). However, this extension from the null field to the magnetically dominated field is possible only for the $a = 0$ case\cite{2013CQGra..30s5012B}. 
To avoid the current sheet on the equator, we select angular functions $P(\theta)$ and $Q(\theta)$ as
\begin{equation}
P=\frac{1}{4}\sin^4 \theta,
~~
Q=\cos \theta \sin^2 \theta.
\label{angsmooth}
\end{equation}
For this model, we have $Q(\pi/2) = 0$, and therefore, $B_{\hat \phi}(\pi/2) = E_{\hat \theta}(\pi/2) = 0$. The discrete function $(1-2H_{\pi/2})$ in eq.~(\ref{selq0m}) is replaced by a continuous function $\cos\theta$ in eq.~(\ref{angsmooth}). There is no singular behavior in the current and charge density on the equator, 
\begin{equation}
4 \pi \rho_{\rm e} =-\frac{r^2+a^2}{\alpha \Sigma^2}
\left[ a^2 W^{\prime \prime} P + W(
3\cos^2\theta-1) 
\right].
\label{chrgQ0}
\end{equation}
The charge density distribution for $a = 0$ is determined by $\theta$ only; it is negative for polar regions ($|\cos\theta| >3^{-1/2}$) and positive for an equatorial region ($|\cos\theta| <3^{-1/2}$), if $W>0$.
%

\section{Particle acceleration}
\subsection{Model}
Some electromagnetic fields are approximated by the null electric current. 
For example, ${\VEC j}=\rho_{\rm e} {\VEC e}_{\hat r}$ in the split-monopole solution. 
The solution is useful to describe a radiative field in the far region. 
Our concern is a null electromagnetic field near a black hole. 
We assume that an electromagnetic burst is temporally generated there, and that the impulsive field is approximated by eq.~(\ref{BfldM0})-(\ref{EfldM0}). 
The consequence for plasma in strong gravity region is studied. 
To model the burst profile, we specify the function $W(u)$ in eq.(\ref{STaxym}) as a function of outgoing time $u$,
\begin{equation}
W(u)= B_{0} M {\rm sech}^2 [(u -u_{0})/\tau_{d}],
%
\end{equation}
where $B_{0}$ denotes a constant representing the field strength, 
$u_{0}$ denotes a constant for the burst peak-time, and the duration of the burst is $\sim \tau_{d}$; further, we fix the timescale as $\tau_{d}=M$. 
In addition, we adopt the angular functions $P(\theta)$ and $Q(\theta)$ in 
eq.(\ref{angsmooth}), which are smooth on the equator.

We consider a charged-particle motion when the impulsive field passes. The equation of motion is given by
\begin{equation}
\frac{dU^{\alpha}}{d\tau}
+\Gamma^{\alpha} _{\beta \gamma} U^{\beta} U^{\gamma}
 =\frac{q}{m}F^{\alpha \beta} U_{\beta},
\label{chargedPEOM}
\end{equation}
where $m$, $q$, and $U^{\alpha}$ denote the mass, charge, and the four-velocity of the particle, respectively. Further, $\Gamma^{\alpha} _{\beta \gamma}$ is the Christoffel symbol and $ F^{\alpha \beta}$ is the Faraday tensor. Equivalently, it may be convenient to derive the equation of motion
from the Lagrangian

\begin{equation}
 L=\frac{m}{2} g_{\alpha \beta}U^{\alpha}U^{\beta} +qA_{\alpha} U^{\alpha}.
%
\end{equation}

Equation (\ref{chargedPEOM}) involves a one-dimensionless parameter 
$\kappa\equiv qB_{0} M/m$. We estimate it for a typical SMBH with $M\sim 10^8M_\odot$: $|\kappa|\approx 10^{13}(B_{0}/{\rm{kG}})$ $(M/10^8M_\odot)$ for an electron, and $|\kappa|\approx 10^{10}(B_{0}/{\rm{kG}})$ $(M/10^8M_\odot)$ for a proton.
This implies that the impulsive field appears as a large amplitude electromagnetic field. The response to the incident field is considerably different depending on $\kappa$. When the particle motion is an oscillatory motion for the small $|\kappa|$, it is accelerated for the large $|\kappa|$. A similar field with $|\kappa|\gg 1$ is produced by a rotating pulsar, and the particle acceleration was discussed in \cite{1969PhRvL..22..728G,1969ApJ...157.1395O,1971ApJ...165..523G}. We focus on exploring a similar acceleration under a strong gravity.
%

\subsection{Results}

\begin{figure}[t]
\centering
 \includegraphics[scale=0.5]{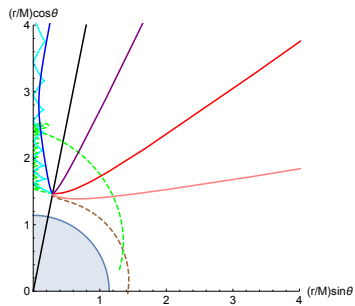} 
\caption{
\label{Fig2}
Examples of trajectories in $(r, \theta)$ plane near the Kerr black hole with $a = 0.99 M$. All trajectories start from the initial position  $(r_{0}, \theta_{0})=$$(1.5M, \pi/16)$; however, they depend on $\kappa$.
The black straight line represents $\theta=\theta_{0}$, and the blue quarter circle represents the horizon. }
\end{figure}

\begin{figure}[t]
\centering
 \includegraphics[scale=0.65]{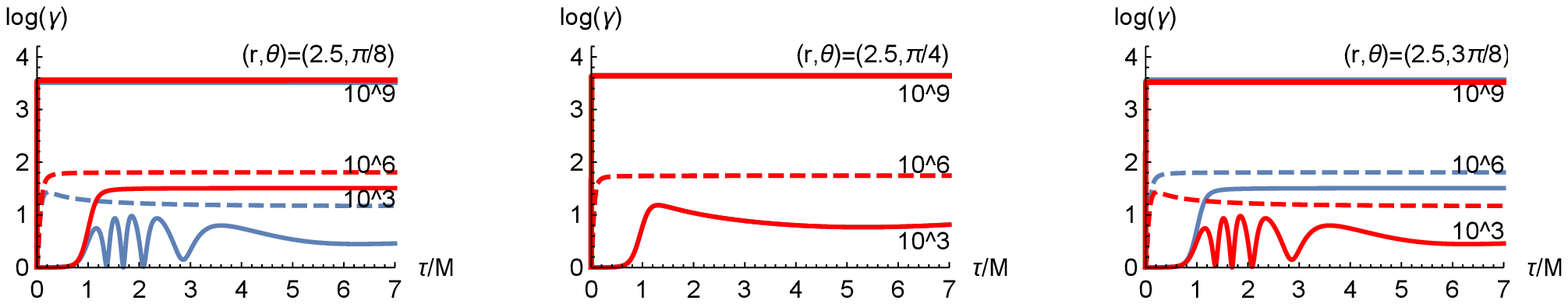}
 \includegraphics[scale=0.65]{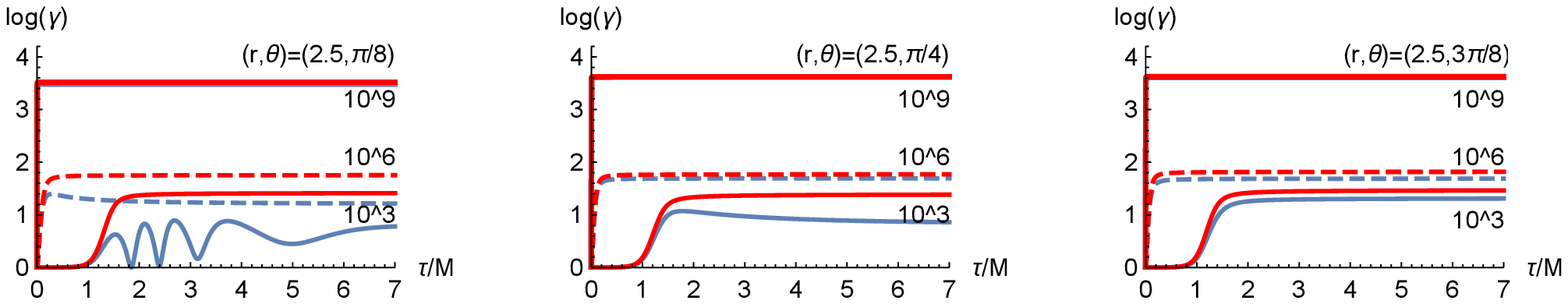}
 \includegraphics[scale=0.65]{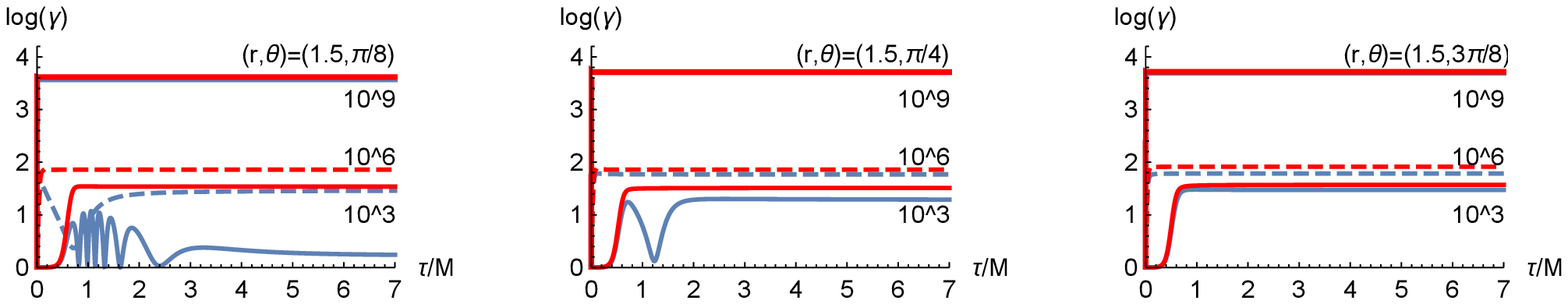}
\caption{
\label{Fig3}
Lorentz factor $\log_{10}(\gamma)$ as a function of proper time $\tau$.
Results are shown for the initial radius $r_{0}=2.5M$ in a Schwarzschild black hole (top panels), for $r_{0}=2.5M$ in a Kerr with $a=0.99M$ (middle panels)and for $r_{0}=1.5M$ and $a=0.99M$ (bottom panels). Blue curves are for positive $\kappa$, whereas red curves denote negative $\kappa$.
}
\end{figure}

We numerically follow the particle's trajectory for a fixed value of $\kappa$. Figure \ref{Fig2} demonstrates some trajectories near a Kerr black hole with $a=0.99M$. A particle starts from the initial position $(r_{0}, \theta_{0})=(1.5M, \pi/16)$, when the burst field passes.
The initial $\theta$-motion depends on the sign of $\kappa$: $U_{\hat \theta}< 0$ for positively charged particle, whereas $U_{\hat \theta}> 0$ for negatively charged particle because $E_{\hat \theta}<0$ in the upper hemisphere. 
For a small value of $|\kappa| \ll 10^3$, the acceleration is very weak.
After the burst, the particle eventually falls owing to the gravitation pull of the central back hole. As $|\kappa|$ increases, the acceleration increases, thereby allowing the particle to escape. The outward direction approaches a line with $\theta =\theta_{0}$ as $|\kappa|$ increases.
We found that it is numerically difficult to simulate the motion of the particle with a considerably large value such as  $>10^{10}$ because of the stiff nature of the differential equation (\ref{chargedPEOM}).
It might be interesting whether to escape near a threshold $\kappa$; however, we focus on particle acceleration for a sufficiently large value of $|\kappa|$. Hereafter, we set $\kappa =\pm 10^3$, $\pm 10^6$, and $\pm 10^9$. Figure \ref{Fig3} shows the evolution of the Lorentz factor 
$\gamma$ as measured by FIDO. 
The top panels from the left to the right show the results for the initial position $(r_{0},\theta_{0})=$$(2.5M, \pi/8)$, $(2.5M, \pi/4)$, and
$(2.5M, 3\pi/8)$ in the Schwarzschild spacetime.
For the strong case $|\kappa| =10^6$, or $10^{9}$, the particle is accelerated in a moment, whereas the particle is alternatively accelerated or decelerated for $|\kappa|=10^3$. This behavior originates from bounces of the $\theta$-motion.

A symmetric property can be seen with respect to the initial angle $\theta_{0}$. For example, the behavior of positively charged particles started from $\theta_{0}=\pi/8$ is the same as that of negatively charged particle started from $\theta_{0}=3\pi/8$.
The electromagnetic field with the angular functions, eq.~(\ref{angsmooth}) 
in the Schwarzschild spacetime, is symmetric with respect to $\theta=\pi/4$.
To study the effect of the spacetime, the results for the Kerr black hole with $a = 0.99M$ are shown in the second row of Fig.~\ref{Fig3}. The initial positions of the particle are the same as those in the first row. The results are not considerably different from those for $a=0$, except for the fact that the symmetry with respect to $\theta=\pi/4$ is no longer observed.

The bottom panels of Fig.~\ref{Fig3} show the results for which the initial position is closer to the black hole, i.e., $r_{0}=1.5M$. The oscillation in $\gamma$ becomes prominent for a particle with $|\kappa|=10^3$ from the polar region, and the terminal Lorentz factor is reduced. This is because gravity becomes stronger by inwardly shifting the initial position.
For accelerated trajectories with $|\kappa|=10^3,10^9$, there is no clear difference between the results with $r_{0}=2.5M$ and those with 
$r_{0}=1.5M$ because the electromagnetic force dominates.
Note that the initial position $(r_{0},\theta_{0})=(1.5M, \pi/8)$ in the left bottom panel of Fig.~\ref{Fig3} is outside the ergo-sphere, whereas others---$(r_{0},\theta_{0})=(1.5M, \pi/4)$ and $(1.5M, 3\pi/8)$---are 
inside the ergo-sphere. The acceleration is the same whether the initial position is the interior or the exterior of the ergo-sphere.
%

\begin{figure}[t]
\centering
 \includegraphics[scale=0.5]{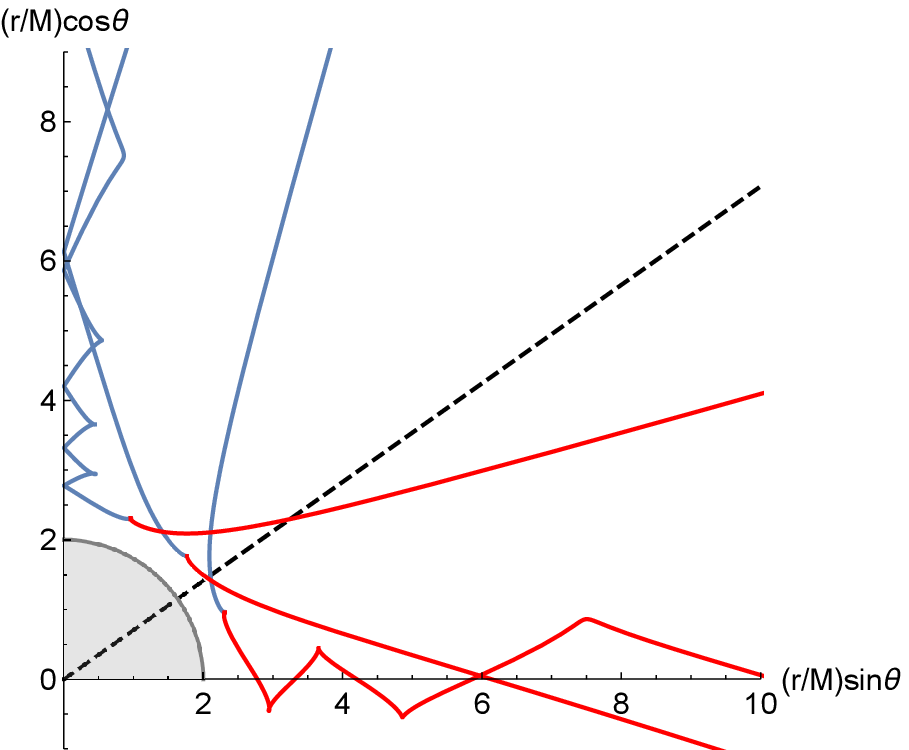}
\hspace{10mm}
 \includegraphics[scale=0.5]{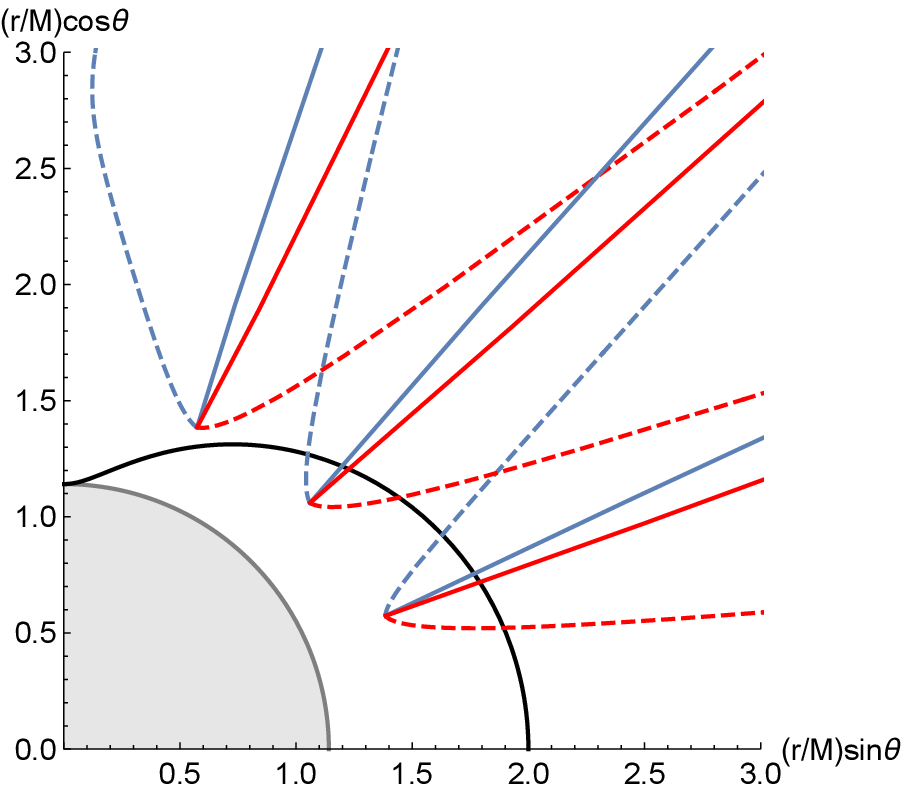}
\caption{
\label{Fig4}
Examples of trajectories in the $(r, \theta)$ plane. The trajectory of the positively charged particles is shown by the blue curve, whereas that of negatively charged particle is shown by the red curve. The left panel shows the results for $|\kappa|=10^3$ in Schwarzschild spacetime. The initial positions are $(r_{0}, \theta_{0})=$$(2.5M, \pi/8)$, $(2.5M, \pi/4)$, and $(2.5M, 3\pi/8)$. The black dotted line represents $\cos\theta=3^{-1/2}$. 
The right panel shows results for $|\kappa|=10^6$ (dotted curve) and $10^9$ (solid curve) in the Kerr spacetime with $a = 0.99M$. The initial positions are $(r_{0}, \theta_{0})=$$(1.5M, \pi/8)$, $(1.5M, \pi/4)$, and $(1.5M, 3\pi/8)$. The evolution of the Lorentz factor for these parameters are shown in the bottom panels in Fig.~\ref{Fig3}. The horizon and ergo-region are shown by the black curve.
}
\end{figure}

Figure \ref{Fig4} shows the particle trajectory in the meridian plane. The initial direction of $\theta$-motion is determined by the sign of $\kappa$; for example, the positively charged particle is accelerated towards the polar region because of $E_{\hat \theta}<0$ in $0< \theta <\pi/2$. In the case of $\kappa=10^3$, a positively charged particle that started from $\theta_{0}=\pi/8$ is reflected on the pole, as shown in the left panel of Fig.~\ref{Fig4}. After the direction of $\theta$-motion changed, the particle is decelerated and the velocity eventually becomes zero. Again, it accelerates towards the polar region.
A negatively charged particle is accelerated towards the equator because of $E_{\hat \theta}<0$ in the region of $0< \theta <\pi/2$. This particle may pass through $\theta =\pi/2$; however, it is decelerated because of $E_{\hat \theta}>0$ in the region of $\pi/2< \theta <\pi$. Thus, the particle goes back to the upper hemisphere.
For positively charged particles started from a near pole or negatively charged particles started from a near equator, their motions are significantly affected by the boundary at $\theta=0$ or $\pi/2$. Thus, the particles undergo deceleration during the burst. The total acceleration is less effective for the small $|\kappa|$ case. Further, we demonstrate the particle trajectories started near the Kerr black hole for large cases $|\kappa| =10^6$ and $10^9$ in the right panel of Fig.~\ref{Fig4}. The acceleration is instantaneous, and therefore, it is not affected by the boundary. The change in angle $\theta$ becomes smaller as 
$|\kappa|$ increases.
As shown in Fig.~\ref{Fig4}, the incident electromagnetic field affects the $\theta$-motion in the opposite direction based on charge in the particle; hence,  it will result in some distribution of the charge density and electric current with respect to the angle $\theta$.
We consider the relation with those in the background field. In the Schwarzschild case, both the sign of the charge density and the direction of the outflow current are distinguished by only $\theta$: $\rho_{\rm e}<0$ in the range of $0 \le \theta < \arccos(3^{-1/2})$, whereas $\rho_{\rm e}> 0$ in the range of $\arccos(3^{-1/2}) < \theta <\arccos(-3^{-1/2})$ (see eq. (\ref{chrgQ0}).) In the left panel of Fig.~\ref{Fig4}, a line of $\theta=\arccos(3^{-1/2})$ is plotted. Figure~\ref{Fig4} shows that the positive particles move toward the pole and the negative ones move toward the equator. This tendency appears in the counteracting direction of the background current flow.
The counteracting flow found by our test-particle approximation means that
the incident field decays by the energy transfer to plenty of particles
 in a realistic model.
Critical number density is $n_{c} \sim j/q$.
This backreaction to the burst field  
is important for $n > n_{c}$, but is beyond present consideration.

\section{Discussion}
Acceleration mechanisms for achieving cosmic-ray energies of
$10^{20}$ eV have been discussed so far.
The mechanisms are basically classified into two types, statistical or direct acceleration \cite{1984ARA&A..22..425H}.
Stochastic acceleration successfully explains power law spectrum, but has the disadvantage of being slow.
On the other hand, the direct process is fast, but does not clearly explain the spectrum.
Particles may be accelerated directly to very high energy by an extended electric field, which is generally associated with the rapid rotation of magnetized compact objects.

Our calculation of charged particle motion driven by null electromagnetic field is also relevant to the direct acceleration mechanism.
We examined the burst field under strong gravity near a black hole.
The typical duration $\tau \sim M$ of the burst is short from a macroscopic view; however, it is sufficiently long for a particle to be accelerated to the relativistic energy. Thus, the interaction between the particle and the burst field is characterized by a large dimensionless number $\kappa = qB_{0}\tau/m \sim qB_{0}M/m$. 
This large number is an essential quantity to determine the maximum energy in direct acceleration mechanisms.
The maximum energy also increases with $\kappa$ in
other acceleration processes near a Kerr black hole, such as magnetic Penrose process (e.g.,
\cite{2019PhRvL.122c5101P,2020ApJ...895...14T}), or wake-field acceleration by intensive Alfvén pulses from an accretion disk
\cite{2018MNRAS.479.2534M}.

Acceleration mechanism
in the radiative zone of a pulsar is discussed in
\cite{1969PhRvL..22..728G,1969ApJ...157.1395O,1971ApJ...165..523G}.
For a wave with a very large amplitude and very low frequency, a particle is instantaneously accelerated to the relativistic regime, and it moves in the direction of wave propagation. 
The null field always corresponds to the radiative zone, and it similarly affects the particle for $|\kappa| \gg 1$. The acceleration becomes more effective with an increase in $|\kappa|$. 
An increase of the Lorentz factor $\gamma$ was calculated by analytic approximation in flat spacetime in previous works
\cite{1969PhRvL..22..728G,1969ApJ...157.1395O,1971ApJ...165..523G}.
Here we performed numerical calculations to obtain an empirical relationship between the maximum Lorentz factor $\gamma_{\rm max}$ and $\kappa~(\gg 1)$, given by
\begin{equation*}
\gamma_{\rm max} \approx 10^{-2} \times \kappa^{2/3}. 
\end{equation*}
We adopted the power index 2/3 of the analytic approximation and obtained numerical coefficient $10^{-2}$. 
The threshold for the acceleration is approximately given by $|\kappa| \sim 10^3$. 
For the values of $|\kappa| \sim 10^3$, the terminal Lorentz factor depends on initial position of the particle and the black hole spacetime. The trajectory is complicated near a critical value, using which the dominant force---electromagnetic acceleration or gravity---is determined. When the acceleration is insufficient, the particle is eventually absorbed in the black hole. However, the gravity is less sensitive for large $|\kappa|$ because the particle is instantaneously accelerated by the very strong electromagnetic field. In the extreme cases of $|\kappa| \gg 1$, the direction of motion is along the propagation of the incident null field. The particle rides the background field at an essentially constant phase.

We estimate the maximum Lorentz factor  $\gamma_{\rm max} $
for a typical SMBH, $\gamma_{\rm max} \sim 10^{5} $
 ($E_{\rm max} \sim 100 $TeV) for a proton. There is ambiguity concerning the field strength $B_{0}$ at various astrophysical sites. However, the upper limit $B_{\rm max}$ is given by $B_{\rm max} \sim M^{-1}$ for a gravitational bound object with mass $M$, since the magnetic energy $B^2M^3$ is less than the rest mass energy $M$.
Thus, the parameter is limited to $\kappa_{\rm max}\approx q/m$, and the maximum energy for a proton $E_{\rm max} \sim 10^{19}$eV is less than the observed highest energy of the cosmic rays.

For a significantly large $|\kappa|$, the acceleration process is instantaneous and is therefore local. Namely, the curved spacetime effect is unimportant.
We consider the burst ejection as a null field to retain the magnetic dominance.
The propagation range may not be very long when the energy transfer to the particle is effective.
However, there are numerous astrophysical situations
where $|\kappa|$ is not extremely large, but moderately large,
e.g., $|\kappa|\sim 10^{3}$.
Some factors in more realistic cases reduce $\kappa$; the timescale $\tau$ associated with local irregularity becomes smaller, and the electromagnetic field-amplitude of a dynamical event is smaller than that of the stationary field, which is inferred from observations.
In scenarios with moderately large $\kappa$, the particle acceleration is complicated, but such a scenario could be interesting to study in the future.
Moreover, the black hole spacetime would be also relevant in such cases.

Finally, we discuss the magnetic field configuration in astrophysical situations.
External magnetic field, e.g., asymptotically uniform field described in 
Fig.~\ref{Fig1}, is likely to coexist with the transient null field considered in this paper.
Such a global magnetic field along the polar axis significantly affects the particle motion.
The trajectory in the transverse plane is circular with a Larmor radius, and particles are confined.
On the other hand, particles along the polar axis are allowed to escape on a straight line.
Thus, the global magnetic configuration determines the outflow structure.
The particle propagation is also an important study in the future.

\begin{acknowledgments}
 This work was supported by JSPS KAKENHI Grant Number 
JP17H06361 and JP19K03850.
\end{acknowledgments} 
%
%

 \end{document}